\documentclass[12pt]{iopart}
\usepackage{graphicx}
\usepackage{bm}
\usepackage{epstopdf}
\newcommand{\eqref}[1]{(\ref{#1})}

\begin{document}
\review{
Spin-dependent phenomena in semiconductors in strong electric fields }

\author{L.~E.~Golub and E.~L.~Ivchenko}
\ead{golub@coherent.ioffe.ru}
\address{Ioffe
Physical-Technical Institute of the Russian Academy of Sciences, 194021
St.~Petersburg, Russia}

\begin{abstract}
We develop a theory of spin-dependent phenomena in the streaming regime characterized by ballistic acceleration of electrons in the moderate electric field until they achieve the optical phonon energy and abruptly emit the phonons. It is shown that the Dyakonov-Perel spin relaxation is drastically modified in this regime, the current-induced spin orientation remarkably increases, reaches a high value $\approx 2$~\% in the electric field $\sim 1$~kV/cm and falls with the further increase in the field.
The spin polarization enhancement is caused by squeezing of the electron momentum distribution in the direction of drift. We also predict field-induced oscillatory dynamics of spin polarization of the photocarriers excited into the conduction band by a short circularly-polarized optical pulse.
\end{abstract}



\maketitle

\section{Introduction}\label{sec:intro}

The orientation of electronic spins in semiconductors by electrical means along with
the possibility to convert the electron spin polarization into an electric signal
are the focus of active research and motivate studies in several directions. A lot of efforts
are aimed at realizing the electrical spin injection from a ferromagnetic metal contact into a semiconductor~\cite{inj1,inj2,inj3}. The second field of research concerns mutual transformation of charge and spin currents and the spin accumulation at the sample edges~\cite{SG_EL}. The third possibility of coupling between spin and electric current, without magnetic materials and role of the sample edges, arises in noncentrosymmetric systems allowing the spin-orbit interaction linear in the electron wave vector~\cite{Rashba_term}. The effect of spin orientation of free carriers by a passage of electric current was first predicted for gyrotropic crystals~\cite{IvchPikus} and observed in bulk tellurium as a current-induced optical activity~\cite{cpge2}, see also \cite{cpge3}, and strained InGaAs epitaxial layers \cite{Kato04}. The theory was extended on two-dimensional (2D) systems lacking a center of inversion~\cite{Vasko,Aronov,Edelstein,Aronov91} and experimentally proved~\cite{Ganichev06,Silov2004}, see also Chap.~9 in the book~\cite{SG_EL}, Ref.~\cite{spinorient_PRB} and references therein. Summarizing the theoretical consideration, the spin polarization $s$ (per particle) created by the electric-current flow can in general be estimated as 
\begin{equation}
\label{s_low_field}
	s =  c \frac{\beta_{\rm so}}{v} \frac{p_{\rm dr}}{p}\:,
\end{equation}
where $c$ is a dimensionless coefficient of the order of unity, $v$ and $p$ are the root-mean-squares of the group velocity and the quasimomentum referred to the extremum point in the Brillouin zone, $p_{\rm dr}$ is the electron drift momentum in the electric field ${\mathcal E}$, and $\beta_{\rm so}$ is a coefficient (in units of velocity) relating the spin-orbit energy with the electron quasimomentum. A value of the induced spin $s$ can be increased by increasing the ratios $\beta_{\rm so}/v$ and (or) 
$p_{\rm dr} / p$. The first ratio is large and has an order of unity in strongly spin-orbit coupled systems, like $p$-type bulk Te, the (111) surface of topological insulator Bi$_2$Se$_3$~\cite{spinorient_PRB,Hosur,Japan_TI} and new classes of noncentrosymmetric systems that emerged recently~\cite{Rashba}. However, in this case the spin is tightly bound to the quasimomentum and loses the degree of freedom. 

\begin{figure}[t]
\includegraphics[width=0.65\linewidth]{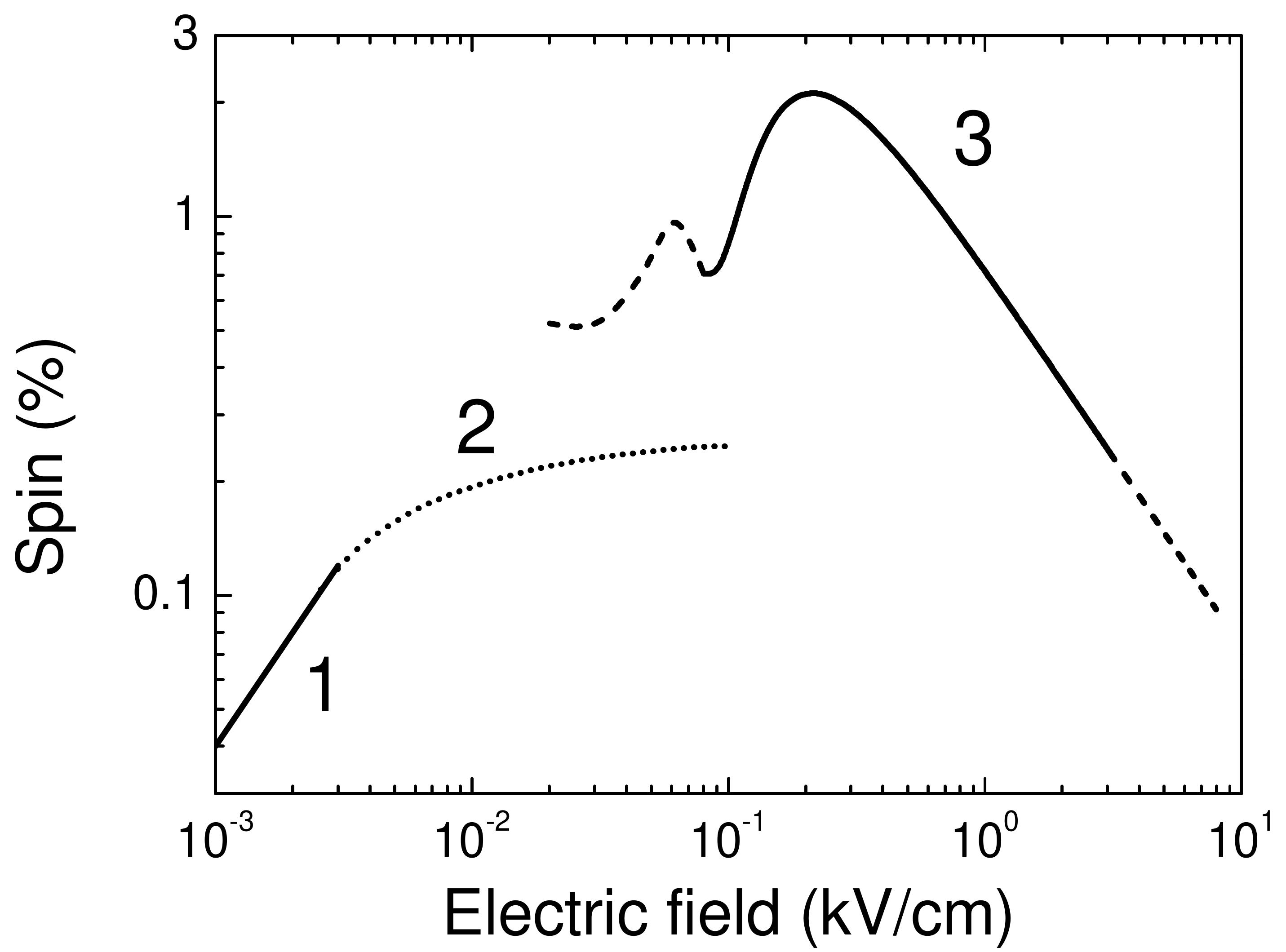}
\caption{Dependence of the induced spin on the electric field. Curve~1 is calculated in the limit of weak fields, Eq.~\eqref{s_low_field}; the dotted curve~2 is an intuitive extrapolation of Eq.~\eqref{s_low_field} for the increasing electric field; curve 3 presents the result of calculation in the streaming regime.}
\label{fig_spin_vs_E}
\end{figure}

In conventional semiconductor 2D systems with a weak spin-orbit coupling, like GaAs- or InAs-based heterostructures, and for comparatively weak fields $\mathcal{E} \ll 1$~kV/cm, each of the two ratios are smaller than $0.1$ resulting in a value of $s \sim 
0.1\%$. This linear dependence of the induced spin on the electric field is presented by curve~1 in Fig.~\ref{fig_spin_vs_E} for 
a degenerate electron gas with the typical parameters: $\hbar\beta_{\rm so}=7$~meV$\cdot$\AA, the transport scattering time $p_{\rm dr} /e\mathcal{E} =10^{- 11 }$~s, the electron effective mass $m=0.07m_0$ ($m_0$ is the free-electron mass),
and 2D concentration $N=10^{11}$~cm$^{-2}$.
In such a system the current-induced spin can be increased only via the second ratio, $p_{\rm dr} / p$. An application of the increasing field leads to a transformation from the linear dependence of drift momentum upon ${\mathcal E}$ to saturation.
If the interparticle collisions play a minor role in the kinetics, the electron momentum-space distribution in strong electric fields becomes extremely anisotropic, or as they say, \textit{streaming}-like. Each electron accelerates quasiballistically in the ``passive'' region until reaching the optical-phonon energy $\hbar \omega_0$ and the quasimomentum $p_0 = \sqrt{2 m \hbar 
\omega_0}$.
Then it loses its energy by emitting an optical phonon and starts the next period of acceleration. Neglecting spin effects, the streaming distribution has been analyzed for the three-dimensional (3D) plasma \cite{Streaming_bulk_review,Andronov80,Dmitriev_Tsendin} as well as for 2D electron gas~\cite{ridley,Streaming_theory_1a,Streaming_theory_1b,Streaming_theory_2}. It is obvious that in this case the ratio $p_{\rm dr} / p$ lies in the range of 0.5. A naive extrapolation of Eq.~\eqref{s_low_field} for the moderate electric fields is shown by the dashed curve 2 in Fig.~\ref{fig_spin_vs_E}. However, in this work we argue that
the spin  reaches a  value remarkably larger than $\sim \beta_{\rm so} m/p$ given by Eq.~\eqref{s_low_field} for $p \sim p_0$. We will show that, actually, in this case the generated spin can be estimated by
\begin{equation} \label{s_high_field}
	s = - \frac{\beta_{\rm so} m}{{6} e \mathcal{E} \tau}\:,
\end{equation}
where $e$ is an electron charge and $\tau$ is the optical phonon emission time  in the ``active'' energy region.
An explicit behavior of the spin orientation in the streaming regime is shown by curve~3  in Fig.~\ref{fig_spin_vs_E} for 
the same value $\hbar\beta_{\rm so}=7$~meV$\cdot$\AA. One can see that the spin polarization is saturated at the high value $s \sim 2\%$ in the fields $\mathcal{E} \geq 0.1$~kV/cm. At the higher fields the spin slightly decreases.

In the present work we develop a theory of spin orientation by the electric current in the streaming regime. Simultaneously we theoretically study relaxation and dynamics of the electronic spin polarization in this regime. 
The paper is organized as follows. In Sec.~\ref{sec:distr}, we summarize the spin-independent theory of the streaming-like transport. In Sec.~\ref{sec:DP}, we propose a theory of spin-dependent high-field transport. Sections~\ref{Sec:el_spin_orient} and~\ref{Sec:strong} are devoted to the spin orientation by electric field.  In Sec.~\ref{Sec:Disc} we discuss the related aspects of the spin-dependent streaming and conclude the paper.

\section{Spin-independent streaming regime}
\label{sec:distr}

In the streaming regime realized in an appropriate range of the dc electric field, the electron distribution in the momentum space has a steaming-like, or needle-like, form spread between the $\Gamma$-point ${\bm p}=0$ and the point ${\bm p}_0 = p_0 \hat{\bm e}$, where $\hat{\bm e}$ is the unit vector in the direction opposite to the electric field $\bm{\mathcal{E}}$. The formation of such an anisotropic electron distribution requires the following hierarchy of times
\begin{equation} \label{tauee}
\tau \ll t_{\rm tr} \ll \tau_p, \, \tau_{\rm ee}\:.
\end{equation}
Here $\tau$ is the time of optical phonon emission by an electron in the active energy region, $t_{\rm tr} = p_0/|e \mathcal{E}|$ is the travel time through the passive region of the momentum space, $\tau_p$ is the momentum relaxation time due to electron scattering by acoustic phonons or static imperfections, and $\tau_{\rm ee}$ is the electron-electron collision time. The electron-electron scattering tends to convert the anisotropic distribution to the shifted
Maxwellian distribution with an effective electron temperature and a drift velocity $p_{\rm dr}/m$. The corresponding time can be estimated by $\tau^{-1}_{\rm ee} \sim (e^2/\kappa)^2 N/(\hbar^2 \omega_0)$, where $N$ is the electron 2D density and $\kappa$ is the dielectric constant. 

Before proceeding with the spin orientation by the electric current in a strong electric field we remind the main steps in description of the spin-independent streaming distribution~\cite{Andronov80}.
We assume the electric field $\bm{\mathcal{E}}$ to be directed antiparallel to the in-plane axis $x$ and satisfy the inequalities~\eqref{tauee}. The two regions are selected in the momentum space, namely, passive with $p \equiv |{\bm p}| < p_0$ and active with $p > p_0$ and positive $p_x$. The electron distribution function $f_{\bm p}$ defined within the passive and active regions is indicated by $f_{\bm p}^p$ and $f_{\bm p}^a$, respectively. 
The kinetic equation for $f_{\bm p}$ breaks up into the two equations
\begin{equation}
	\label{kinetic_eq}
	e \mathcal{E} \frac{\partial f_{\bm p}^a}{\partial p_x} = - {f_{\bm p}^a \over \tau}\:,\qquad
e \mathcal{E} {\partial f_{\bm p}^{p}\over \partial p_x} = \sum_{\bm p'} W_{\bm p \bm p'} f_{\bm p'}^a	\:.
\end{equation}
Here $W_{\bm p {\bm p}'}$ is the probability rate for the longitudinal-optical (LO) phonon emission in the electron transition 
${\bm p}' \to {\bm p}$, with $p' > p_0$; it is given by Fermi's golden rule
\begin{equation} 
\label{Q}
W_{\bm p {\bm p}'} = {2\pi\over \hbar} \sum_{q_z} \left|M_{\bm p {\bm p}'}(q_z)\right|^2 \delta\left(E_p - E_{p'} + \hbar\omega_0\right) \:,
\end{equation}
where $E_p$ is the electron energy $p^2/2m$, $\tau$ is the inverse total emission rate $\sum\limits_{{\bm p}} W_{\bm p \bm p'}$, it also enters the inequality~\eqref{tauee}, $M_{\bm p {\bm p}'}(q_z)$ is the matrix element for emission of the LO phonon with the 
in-plane wave vector ${\bm q}_{\parallel} = ({\bm p}' - {\bm p})/\hbar$ and the vertical component $q_z$. In bulk 
zinc-blende-lattice semiconductors, one has for the Fr\"ohlich electron-phonon interaction
$M^{({\rm 3D})}_{{\bm p} {\bm p}^{\prime}}= {\rm i} \hbar C_{\rm F}/\left(\sqrt{V} |{\bm p} - {\bm p}'|\right)$,
where $C_{\rm F}$ is the polar-optical interaction constant~\cite{cardona}, $V$ is the crystal volume and ${\bm p}, {\bm p}'$ are the 3D electron quasimomenta. In a quantum well, the Fr\"ohlich coupling is renormalized as follows
\begin{equation} \label{froehlich2}
M_{{\bm p} {\bm p}^{\prime}}(q_z)= \frac{{\rm i} C_{\rm F}}{\sqrt{V \left(q_\parallel^2 + q_z^2\right)}} 
\int\limits_{-\infty}^\infty dz {\rm e}^{{\rm i} q_z z} \varphi^2(z) \:,
\end{equation}
where $\varphi(z)$ is the electron size-quantized envelope wavefunction. After the summation over $q_z$ one obtains
\begin{equation}
\label{tausc}
{1\over \tau} = \frac{m C_{\rm F}^2}{2\hbar^3 q_{\parallel} } \int\limits_{-\infty}^\infty dz \int\limits_{-\infty}^\infty dz' {\rm e}^{-q_{\parallel} |z-z'|} \varphi^2(z) \varphi^2(z')\:.
\end{equation}

The two branches of the distribution function found by solving the first and second equations (\ref{kinetic_eq}) are 
sewed at the circle $|{\bm p}|=p_0$ 
by the continuity boundary condition $ f_{{\bm p}}^p \vert_{p=p_0} = f_{{\bm p}}^a \vert_{p=p_0}$. 
The solution normalized to the 2D concentration $N$ reduces to~\cite{ridley}
\begin{eqnarray}
\label{spin_independ}
f_{\bm p}^p = f_0 {\rm exp}\left( - \frac{\alpha}{2} \frac{p_y^2}{p_0^2}\right) \tilde{\theta} \left( p_x \right)\:,\\
f_{\bm p}^a = f_0 {\rm exp}\left( - \frac{\alpha}{2} \frac{p_y^2}{p_0^2}\right) \, {\rm exp} \left( - \alpha {p_x - p_{0x} \over p_0}\right)\:,\nonumber
\end{eqnarray}
where $\alpha = p_0/|e\mathcal{E}|\tau$, $\tilde{\theta} \left( p_x \right) = \left[1 + {\rm erf} \left(\sqrt{\alpha/2} p_x/ p_0 \right) \right]/2$,
$f_0=N \sqrt{2 \pi\alpha} \, \pi \hbar^2/p_0^2$, and $p_{0x} = \sqrt{p_0^2 - p_y^2}$. Note that, strictly speaking, in the second equation (12) of Ref.~\cite{ridley} the difference ${k_x - k_0 = (p_x - p_0)/\hbar}$ should be replaced by ${k_x - k_{0x} }$ with $k_{0x}$ equal to $\sqrt{k_0^2 - k_y^2}$ rather than merely $k_0$. 
One can see from Eq.~(\ref{spin_independ}) that the penetration $\delta p_x$ into the active region and the width $2 \delta p_y$ of the distribution in the transverse direction $p_y$ are given by $\delta p_x = p_0/\alpha$ and $\delta p_y = p_0/\sqrt{\alpha}$. Thus, the needle-like distribution is formed provided $\alpha \gg 1$ which establishes the upper limit for the electric field strength. In this case the function $\tilde{\theta} \left( p_x \right)$ can be approximated by the Heaviside  function $\theta(p_x)$ and $p_{0x}$ by $p_0 - (p_y^2/2 p_0)$. It is also worth to mention that the relative number of particles in the active and passive regions given by  $N_a/N_p \approx	N_a/N 	\approx \alpha^{-1}$ coincides with the ratio $\tau/t_{\rm tr}$ of the times spend by electrons in these regions.

Let us present estimations of the characteristic times for GaAs-based structures where $m = 0.067m_0$, $\hbar\omega_0 =36$~meV, $p_0/\hbar = 2.8\times 10^6$~cm$^{-1}$, and $\tau = 10^{-13}$~s. In this case the condition $\alpha \gg 1$, or equivalently $p_0/ \tau \gg |e \mathcal{E}| $, is met if $\mathcal{E} \ll 16$ kV/cm.  
On the other hand, the condition $t_{\rm tr} \ll \tau_p$ with a typical value $\tau_p=10^{-11}$~s of the momentum relaxation time  requires $\mathcal{E} >0.1$~kV/cm. In the field $\mathcal{E}=0.5$~kV/cm one has $\alpha = 33$, $\delta p_x / p_0 \sim 0.03$, $\delta p_y / p_0 \sim 0.2$, and the travel time $t_{\rm tr} = \alpha \tau = 3.3 \times 10^{-12}$~s is longer than $\tau$ but shorter than $\tau_p$. For the time $\tau_{ee}$ we obtain an estimation 
$\sim  (2~{\rm ps})/(N/10^{11}~\mbox{cm}^{-2})$ which means that the condition $t_{\rm tr} \ll \tau_{ee}$ is satisfied  at $N < 10^{11}$~cm$^{-2}$ similarly to GaN-based structures~\cite{Streaming_theory_1a}. Another restriction on the density $N$ is imposed from the requirement $f_{\bm p} \ll 1$ which allows to exclude the Pauli blocking of the electron acceleration in the passive region. This requirement is equivalent to the inequality $N \ll N_{\rm max}$, where $N_{\rm max}$ is the number of states in the needle, i.e., in the rectangular area of the 2D quasimomentum space with the sides $p_0$ and $2 \delta p_y$. It is also  met for $N \ll 10^{11}$ cm$^{-2}$. 
Note that in the experiment~\cite{IvAnov} the above conditions are not fulfilled.

\section{Spin-dependent kinetic theory}
\label{sec:DP}

The spin-dependent kinetic theory operates with the electron spin density matrix $\rho_{\bm p}$ describing the electron distribution both in the quasi-momentum and spin spaces. This ${\bm p}$-dependent 2$\times$2 matrix may conveniently be presented as
\begin{equation} \label{rhofs}
\rho_{\bm p} = f_{\bm p} + \bm \sigma \cdot {\bm S}_{\bm p}\:,
\end{equation}
where $\sigma_{\alpha}$ ($\alpha = x,y,z$) are the Pauli spin matrices, the distribution function $f_{\bm p} = {\rm Tr}\{ \rho_{\bm p}\} /2$ is the average occupation of the two spin states with the same ${\bm p}$, and ${\bm S}_{\bm p} = {\rm Tr} \{ \rho_{\bm p} {\bm \sigma}/2 \}$ is the average spin of an electron occupying the point ${\bm p}$ of the quasi-momentum space.
The 2D electron concentration and spin polarization per particle are given by $N = 2 \sum\limits_{\bm p} f_{\bm p}$, ${\bm s} = \sum\limits_{\bm p} {\bm S}_{\bm p}/N$. In addition to the parabolic term ${\bm p}^2/2m$, we include in the effective electron Hamiltonian the spin-dependent contribution 
\begin{equation} \label{H_SO}
H_{\rm so}(\bm p) = \beta_{ij} \sigma_i p_j= 	(\hbar/2) \bm \sigma \cdot \bm \Omega_{\bm p}\:,
\end{equation}
where ${\bm \Omega_{\bm p}}$ is a linear function of ${\bm p}$. 
The role of cubic in $\bm p$ terms is analyzed in Sec.~\ref{Sec:Disc}.
Bearing in mind both symmetric and asymmetric quantum-well structures grown along the [001] crystallographic direction we take ${\bm \Omega_{\bm p}}$ in the form of a 2D pseudovector with the components 
\begin{equation} \label{betabeta}
\Omega_{{\bm p},x} = 2\beta_{xy} p_y/\hbar\:,
\qquad \Omega_{{\bm p},y} = 2 \beta_{yx} p_x/\hbar\:,
\end{equation}
where the in-plane axes $x \parallel [1\bar{1}0]$, $y \parallel [110]$ are used and the coefficients  $\beta_{xy}, \beta_{yx}$ have the units of velocity.

The kinetic equation for the spin density matrix $\rho_{\bm p}$
in the presence of spin-orbit interaction $H_{\rm so}(\bm p)$ 
has been derived in Refs.~\cite{ILGP89,ILGP90} for elastic momentum scattering. Applying the similar approach for the dominant optical-phonon emission regime under study we obtain the kinetic equation in the following form
\begin{equation} \label{spin_kinetic}
e \mathcal{E} {\partial \rho_{\bm p} \over \partial p_x} + {{\rm i}\over \hbar} [H_{\rm so}(\bm p),\rho_{\bm p}]
= {\rm St} \{ \rho_{\bm p} \}\:.
\end{equation}
Taking into account the spin-dependent contribution to the effective Hamiltonian the collision term 
${\rm St} \{ \rho_{\bm p} \}$
reads
\[
 {2\pi\over\hbar} \sum_{\bm p'} \left|M_{\bm p {\bm p}'}\right|^2 
	\biggl[
	\left\{ \rho_{{\bm p'}},	\delta [ E_{p'}+ H_{\rm so}({\bm p'}) - E_{p} - H_{\rm so}({\bm p}) - \hbar\omega_0 ] \right\}_s 	
	- ({\bm p}' \leftrightarrow {\bm p}) 	\biggr].
\]
Here the symbols $[A,B]$ and $\left\{A,B\right\}_s$ mean, respectively, the commutator ${AB - BA}$ and the anticommutator $ (AB+BA)/2$, and the term $({\bm p}' \leftrightarrow {\bm p})$ means the previous one with the interchanged variables ${\bm p}'$ and ${\bm p}$. While deriving the collision term we ignored the spin-flips under the optical phonon emission, their role is discussed in Sec.~\ref{Sec:Disc}. The contribution of spin-orbit Hamiltonians included in the $\delta$-functions to observable effects is determined by the quantum parameter $\beta_{\rm so}/v$ which, in particular, enters Eq.~\eqref{s_low_field}. We take this parameter in the first order of perturbation theory which,
for the linear-${\bm p}$ dependence in Eq.~\eqref{H_SO}, 
allows to approximate $\delta[E_{p'} + H_{\rm so}({\bm p}') - E_{p} - H_{\rm so}({\bm p}) - \hbar \omega_0]$ by
\begin{equation}
	\delta(E_{p'} - E_{p} - \hbar \omega_0) + m \frac{\partial H_{\rm so}(\bm p)}{\partial p_j} \left( \frac{\partial }{\partial p_j} + \frac{\partial }{\partial p'_j} \right) \delta(E_{p'} - E_{p} - \hbar \omega_0) \:.
\end{equation}
We can also use the convenient identity
\[
\left\{ \rho_{\bm p}, \frac{\partial H_{\rm so}(\bm p)}{\partial p_j} \right\}_s 
= \beta_{ij} \left(  \sigma_i f_{\bm p} + S_{{\bm p}i} \right)\:.
\]

Now we can readily perform the anticommutation and reduce the matrix equation~(\ref{spin_kinetic}) to a coupled set of the scalar equation for the distribution function $f_{\bm p}$ and the pseudovector equation for ${\bm S}_{\bm p}$
\[
{\cal I}_{\bm p} \left( f_{\bm p} - \frac{\hbar m}{2} {\partial \Omega_{{\bm p}i} \over \partial p_j} ~ {\partial S_{{\bm p}i} \over \partial p_j} \right) = \frac{\hbar m}{2} {\partial \Omega_{{\bm p}i} \over \partial p_j} 
{\partial ( \bm S_{\bm p} \times \bm \Omega_{\bm p})_i \over \partial p_j}, 
\]
\begin{equation}
\label{kin_eq_S_p}
	{\cal I}_{\bm p} \left( {\bm S}_{\bm p}
		- {\hbar m \over 2} {\partial \bm \Omega_{\bm p} \over \partial p_j}  {\partial f_{\bm p} \over \partial p_j} \right) + {\bm S}_{\bm p} \times {\bm \Omega}_{\bm p} = 0\: .
\end{equation}
Here the spin-independent relaxation operator ${\cal I}_{\bm p}$ is defined by~\cite{noci}
\begin{equation}
\label{I}
	{\cal I}_{\bm p} (\Phi_{\bm p}) = e \mathcal{E} {\partial \Phi_{\bm p} \over \partial p_x} + \sum_{\bm p'} (W_{\bm p' \bm p} \Phi_{\bm p} - W_{\bm p \bm p'} \Phi_{\bm p'})\:.
\end{equation}
Clearly, the kinetic equation~\eqref{kinetic_eq} for the distribution function $f_{\bm p}$ considered in the previous section reduces to ${\cal I}_{\bm p} \left(f_{\bm p}\right)=0$. While deriving
Eqs.~(\ref{kin_eq_S_p}) it is taken into acount that the matrix element of optical-phonon emission depends only on the difference ${\bm p - \bm p'}$ in which case the sum of two derivatives $({\bm \nabla}_{\bm p} + {\bm \nabla}_{{\bm p}'}) |M_{\bm p {\bm p}'}|^2$ vanishes.

\section{Electrical spin orientation}
\label{Sec:el_spin_orient}

In this Section we consider the collision-dominated spin dynamics realized for weak spin-splittings satisfying the condition 
$\Omega_{\bm p} t_{\rm tr} \ll 1$, i.e., for a small angle of spin rotation $\phi$ that occurs during a single passage of the passive region, and ignore the spin rotation in the active region during the shortest time $\tau$, see Eq.~(\ref{tauee}). In this case the characteristic times of spin precession and relaxation as well as the establishment time of field-induced spin polarization are much longer than the travel time $t_{\rm tr}$. This allows one to search the spin distribution ${\bm S}_{\bm p}$ in the form
\begin{equation} \label{expansion}
{\bm S}_{\bm p} =  2 f_{\bm p} {\bm s} + \delta {\bm S}^{(1)}_{\bm p} + \delta {\bm S}^{(2)}_{\bm p}\:,
\end{equation}
where ${\bm s}$ is the average spin per particle independent of ${\bm p}$, and the two other terms are corrections of the first and second order in ${\bm \Omega}_{\bm p}$ vanishing after the summation over ${\bm p}$. 
If the electric field is directed along the $x$ axis the system retains the mirror reflection plane $\sigma_v$ perpendicular to 
$y$. Since the unit 2$\times$2 matrix and the Pauli matrix $\sigma_y$ are invariant and the Pauli matrices $\sigma_x, \sigma_z$ change the sign under the $\sigma_v$ operation, the functions $f_{\bm p}$, $S_{{\bm p},y}$ are even and 
$S_{{\bm p},x}, S_{{\bm p},z}$ are odd with respect to this operation. 
Therefore, despite the average $x$- and $z$-components of the spin corrections vanish, the average spin ${\bm s}$ induced by the current is nonzero and polarized along the $y$ axis.

While considering the time-dependent spin dynamics one should add into Eqs.~(\ref{kin_eq_S_p}) the time derivatives $\partial f_{\bm p}/\partial t$ and $\partial {\bm S}_{\bm p}/\partial t$. We take them into account only on a scale of long times, namely, the time of spin relaxation. Summing the kinetic equation for ${\bm S}_{\bm p}$  over $\bm p$ and dividing the result by the electron concentration $N$ we obtain the balance equation for the nonzero average spin component
\begin{equation} \label{s_DP}
{\partial s_y \over \partial t}  + {1\over N} \sum_{\bm p} (\bm S_{\bm p} \times \bm \Omega_{\bm p})_y = 0\:.
\end{equation}
Substituting the expansion (\ref{expansion}) into the sum we can rearrange Eq.~(\ref{s_DP}) into 
\begin{equation} \label{s_DP2}
{\partial s_y \over \partial t}  + \frac{s_y}{\tau_{sy}} = \frac{G_s}{N} \:.
\end{equation}
Here $G_s$ and $\tau_{sy}$ are the spin generation rate and relaxation time of the $s_y$ component defined by
\begin{eqnarray} \label{Gsy}
&&G_s = - \sum_{\bm p} \left(\delta{\bm S}^{(2)}_{\bm p} \times {\bm \Omega}_{\bm p}\right)_y
= - \sum_{\bm p} \Omega_{{\bm p},x} \delta S^{(2)}_{{\bm p},z} \:,\:\\
&&\frac{s_y}{\tau_{sy}} = \frac{1}{N} \sum_{\bm p} \left(\delta{\bm S}^{(1)}_{\bm p} \times {\bm \Omega}_{\bm p}\right)_y
= \frac{1}{N} \sum_{\bm p} \Omega_{{\bm p},x} \delta S^{(1)}_{{\bm p},z}\:. \label{spintime}
\end{eqnarray}

According to the second equation~(\ref{kin_eq_S_p}) we can present the first-order correction as a sum of 
\begin{equation} \label{firstterm}
\delta {\bm S}^{(1a)}_{\bm p} = {\hbar m \over 2} {\partial \bm \Omega_{\bm p} \over \partial p_j}  {\partial f_{\bm p} \over \partial p_j}
\end{equation}
and the correction $\delta {\bm S}^{(1b)}_{\bm p}$ which satisfies the equation 
\begin{equation} 
e \mathcal{E} \frac{\partial}{\partial p_x} \delta {\bm S}^{(1b)}_{\bm p}  + 2 f_{\bm p} {\bm s} \times \bm \Omega_{\bm p} = 0
 \label{sprime}
\end{equation}
in the passive region. The collision term is neglectred here because it is determined by the correction $\delta {\bm S}^{(1b)}_{\bm p}$ in the active region which has an order of $\Omega_{\bm p} \tau \ll \Omega_{\bm p}t_{\rm tr} \ll 1$. The first term $\delta {\bm S}^{(1a)}_{\bm p} $ makes no contribution to the spin relaxation rate, and we focus on the second term $\delta {\bm S}^{(1b)}_{\bm p}$. 

Let us introduce the 2D vector $\bm k=(k_x,p_y)$ 
with an arbitrary $x$-component $k_x$ and the $y$-component coinciding with that of the vector ${\bm p}$. 
Then the solution of Eq.~(\ref{sprime}) can be presented as
\begin{equation} \label{deltasprime}
\delta S^{(1b)}_{{\bm p},z} = - \frac{2}{e \mathcal{E}} \int\limits_{-\infty}^{p_x} dk_x f_{\bm k} \left( {\bm s} \times {\bm \Omega}_{\bm k} \right)_z = 2 s_y \frac{p_x \Omega_{{\bm p}x}}{e \mathcal{E}} f_0 {\rm exp}\left( - \frac{\alpha}{2} \frac{p_y^2}{p_0^2}\right)\:.
\end{equation}
The substitution of this expression into Eq.~\eqref{spintime} and summation over ${\bm p}$ result in
\begin{equation} \label{tauxyz}
{1\over \tau_{sy}} = 2 \tau \left( \frac{\beta_{xy} p_0}{\hbar}\right)^2\:.
\end{equation}

Now we turn to calculating the spin generation rate $G_s$. For large values of the parameter $\alpha$ it suffices to find the second-order correction only in the passive region where it satisfies the equation
\begin{equation} \label{s2z}
e \mathcal{E} {\partial \delta S^{(2)}_{{\bm p},z} \over \partial p_x} + \left( \delta {\bm S}_{\bm p}^{(1a)}
\times \bm \Omega_{\bm p} \right)_z = 0
\end{equation}
with $\delta{\bm S}_{\bm p}^{(1a)}$ defined by Eq.~(\ref{firstterm}).
The straigtforward integration leads to
\begin{equation} \label{s2}
\delta S^{(2)}_{{\bm p},z}  = \frac{m \hbar}{2 e \mathcal{E}} 
\left( - \Omega_{\bm p,y} {\partial \Omega_{\bm p,x} \over \partial p_y} {\partial f_{\bm p}^p \over \partial p_y} {p_x\over 2} + \Omega_{\bm p,x} {\partial \Omega_{\bm p,y} \over \partial p_x} f_{\bm p}^p \right) 
\end{equation}
which results in
\begin{eqnarray}
G_s =  - {m\over e \mathcal{E}} {\hbar \over 2}  \sum_{\bm p}  \left[ \Omega_{\bm p,y} \left( {\partial \Omega_{\bm p,x} \over \partial p_y} \right)^2 {p_x \over 2} + \Omega_{\bm p,x}^2 {\partial \Omega_{\bm p,y} \over \partial p_x}  \right] f_{\bm p}\:.
\label{generation Gs}
\end{eqnarray}
The second term should be ignored because it has an order of $\alpha^{-1} \ll 1$ as compared to the first one.
Then we have for the spin generation 
\begin{equation} \label{gsfinal}
G_s = - \frac{m \hbar N}{24 e \mathcal{E}}\left( {\partial \Omega_{\bm p,x} \over \partial p_y} \right)^2  
\Omega_{{\bm p}_0,y} p_0 \:.
\end{equation}
In the steady-state conditions, the spin polarization $s_y = {G_s \tau_{sy}/N}$ reduces to
\begin{equation} \label{s_fin}
s_y= -{\alpha \over 6} {m\beta_{yx}\over p_0} = -{m\beta_{yx}\over 6 e \mathcal{E} \tau} \:.
\end{equation}
This result exceeds the value $m\beta_{yx} / p_0$ expected from Eq.~\eqref{s_low_field} by the factor $\alpha/6 \gg 1$. The enhancement of the spin polarization can be readily understood taking into account that the spin generation rate $G_s$ is proportional to $\alpha \tau$ while the spin relaxation time $\tau_{sy} \propto \tau^{-1}$ so that the product $G_s \tau_{sy} \propto \alpha$.

The applied straightforward procedure resulting in Eq.~\eqref{s_fin} has allowed us to obtain the numerical factor $-1/6$, 
trace the sequence of mathematical manipulations and determine criteria for validity of the result. Now we give a qualitative interpretation of the spin orientation by the electric current in the streaming regime bringing to light the nature of this effect. Since the generation rate $G_s$ is a third-order correction in the perturbation theory we should successfully consider three stages of action of the spin-orbit interaction on the electron spin. 

\begin{figure}[t]
\includegraphics[width=\linewidth]{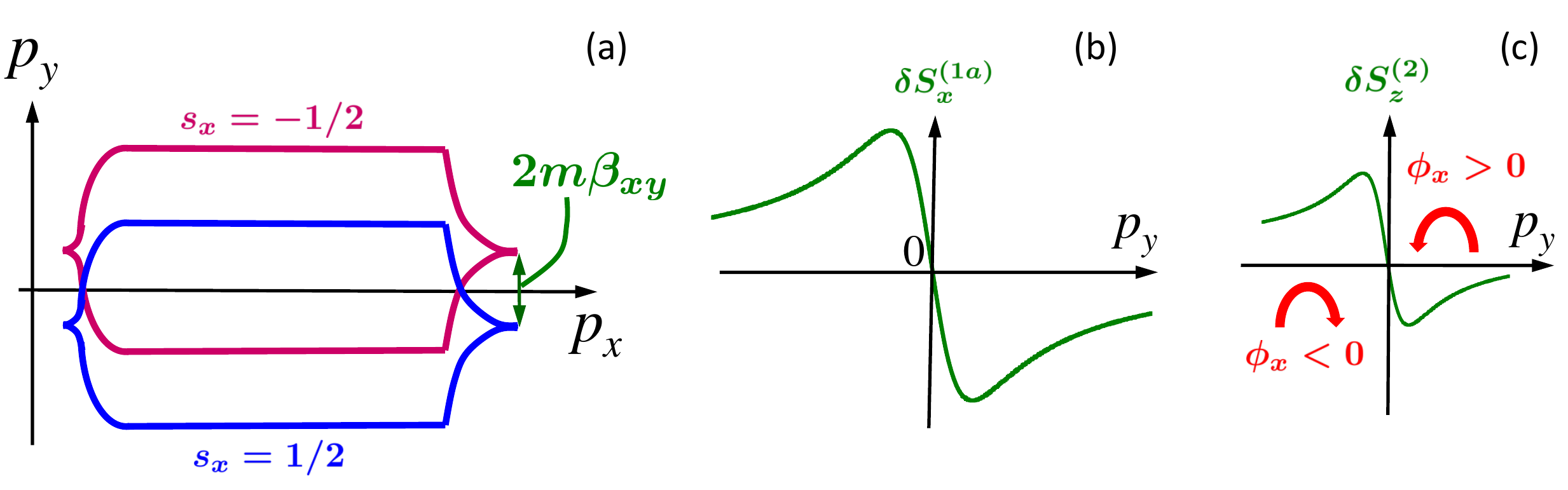}
\caption{Microscopic mechanism of spin orientation in the streaming regime. (a) The distribution functions in the ${\bm p}$ space for electrons with the spin components $s_x = \pm 1/2$ taking into account the spin splitting proportional to 
$\beta_{xy}$. (b) The corresponding spin $x$-component as a function of $p_y$ at a fixed value of $p_x$. (c) The spin polarization along the $z$ axis arising due to the precession around the $y$ axis with the frequency $\Omega_{\bm p,y}$. The arched arrows illustrate the precession of $\delta S_z^{(2)}$ around the $x$ axis with the frequency $\Omega_{\bm p,x}$ yielding the nonzero average spin $\bm s \parallel y$.}
\label{fig_spin_orient_DP}
\end{figure}

At the first stage we make allowance only for the spin-orbit term $(\hbar/2) \Omega_{\bm p,x} \sigma_x = \beta_{xy}p_y \sigma_x$. 
It leads to the splitting of the 2D conduction subband into the two branches $E_{{\bm p}, \pm 1/2} = E_p \pm \beta_{xy}p_y$ for the states with well-defined spin orientation $\pm 1/2$ along the $x$ axis. The ${\bm p}$-dependence of $E_{{\bm p}, \pm 1/2} 
$ can be presented as a 2D parabolic function with the minimum point shifted along the $p_y$ axis by 
$\mp m \beta_{xy}$. Since in this section we neglect any spin-flip relaxation processes the streaming regime is independently established in each branch. The distribution function in the branch $s_x = \pm 1/2$ is obtained from (\ref{spin_independ}) by shifting the argument ${\bm p}$ by the vector $\pm {\bm \kappa}$ with ${\bm \kappa} = (0, m \beta_{xy})$, namely, 
$$f_{{\bm p}, \pm 1/2}^p = f_{{\bm p} \pm {\bm \kappa} }^p \approx f_{\bm p}^p \pm m \beta_{xy} \frac{\partial f_{\bm p}^p}{\partial p_y}\:.$$
The streaming-like 2D electron distributions in the ${\bm p}$ space shifted along $p_y$ axis are schematically shown in 
Fig.~\ref{fig_spin_orient_DP}a for a positive value of $\beta_{xy}$. Thus, in the first-order approximation the electrons are characterized by the spin polarization $\delta S_x^{(1a)} = m \beta_{xy}~ \partial f_{\bm p}^p/ \partial p_y$ coinciding with the correction~(\ref{firstterm}) corresponding to $j = y$; it is depicted in Fig.~\ref{fig_spin_orient_DP}b for $\beta_{xy} > 0$. The correction due to $j=x$, or due to $\Omega_{{\bm p},y}$, leading to the second term in Eq.~(\ref{generation Gs}) can be neglected because as stated above the corresponding contribution to the spin orientation has a small factor $\alpha^{-1}$.

At the second stage we switch in the precession of the spin $\delta S_x^{(1a)}$ around the axis $y$ with the frequency $\Omega_{{\bm p},y}$. During a single field-driven passage of the electron through the passive region the spin $x$-component rotates by the angle $\phi_{y} \sim \Omega_{{\bm p},y} t_{\rm tr}$ giving rise to the spin $z$-component $\delta S_z^{(2)} \sim \phi_y 
\delta S_x^{(1a)}$. The appearance of this component is described by the first term in Eq.~(\ref{s2}). 

The third stage overlaps with the second one and occurs during the same single passage: the appearing $z$-component of the electron spin is further affected by the procession frequency $\Omega_{{\bm p},x}$ to rotate around the $x$ axis by the angle $\phi_x \sim
\Omega_{{\bm p},x} t_{\rm tr}$, see arched arrows in Fig.~\ref{fig_spin_orient_DP}. As a result, the electron reaches the edge between the passive and active regions getting the spin 
\[
\delta S_y^{(3)} \sim \phi_x \delta S_z^{(2)} \sim \phi_y \phi_x \delta S_x^{(1a)} \sim \phi_y \phi_x m \beta_{xy} \frac{\partial f_{\bm p}^p}{\partial p_y}\:.
\] 
This spin polarization income arises within the travel time $t_{\rm tr}$ which means that the spin generation rate can be estimated by $G_s \sim \delta S_y^{(3)}/t_{\rm tr}$. Expressing $\delta S^{(3)}$ in terms of ${\bm \Omega_{\rm p}}$ and $t_{\rm tr}$ and taking $p_x \sim p_0, p_y \sim p_0/\sqrt{\alpha}$ we finally get an estimation consistent with the analytical results~\eqref{gsfinal} and~\eqref{s_fin}. It should be emphasized that neither $\delta S_x^{(1a)}$ nor $\delta S_z^{(2)}$ yield a non-zero net spin polarization and eventually $\delta S_y^{(3)}$ 
turns out to be nonzero after averaging in the ${\bm p}$ space.

\section{Remarkable spin-orbit splitting}
\label{Sec:strong}

The equation (\ref{s_fin}) for the electron spin induced by the electric current is derived assuming small values of the elementary spin-precession angles $\phi_x$ and $\phi_y$. In the streaming regime this is acceptable for $\phi_x$ because, for $\alpha \gg 1$, the spread of the electron distribution along the $p_y$ axis is narrow and the angle $\Omega_{{\bm p}_0,x} t_{\rm tr}$ is small. As for the angle $\Omega_{{\bm p}_0,y} t_{\rm tr} = 2 \beta_{yx} p_0 t_{\rm tr}/\hbar$, its values can be comparable with or exceed unity. Indeed, in GaAs-based quantum-wells the linear-${\bm p}$ coefficient can be estimated by $\hbar \beta_{yx} = 7$~meV$\cdot$\AA~\cite{fabian} which corresponds to the spin splitting $2 \beta_{yx} p_0=0.4$~meV and $2 \beta_{yx} p_0 t_{\rm tr}/\hbar \approx$ 2 for $t_{\rm tr}$ = 3.3$\times$10$^{-12}$ s. Since $\phi_x$ is small the equations~\eqref{firstterm},~\eqref{sprime} are applicable. 
However the spin correction $\delta\bm S^{(2)}_{\bm p}$ should be calculated for an arbitrary value of 
$\Omega_{{\bm p},y} t_{\rm tr}$. We search for the spin density ${\bm S}_{\bm p}$ in the form (\ref{expansion})
with $\delta \bm S^{(1)}_{\bm p} = m \beta_{xy}(\partial f_{\bm p}/\partial p_y) \hat{\bm e}_x$, where $\hat{\bm e}_x$ is the unit vector in the $x$ direction. According to the general equation~\eqref{kin_eq_S_p} the evolution of $\delta \bm S^{(2)}_{\bm p}$ in the passive region is described by 
\begin{equation} \label{vectoreq}
e \mathcal{E} {\partial \delta \bm S^{(2)}_{\bm p} \over \partial p_x} + \delta \bm S^{(2)}_{\bm p} \times \bm \Omega_{\bm p}  + \delta \bm S^{(1)}_{\bm p} \times \bm \Omega_{\bm p} = 0\:.
\end{equation}
As compared to Eq.~(\ref{s2z}) we take explicitly into account the precession of the vector $\delta \bm S^{(2)}_{\bm p}$. 
In Eq.~(\ref{vectoreq}) we can neglect the small $x$-component of the precession frequency ${\bm \Omega}_{\bm p}$ because it is already contained in Eq.~(\ref{Gsy}).  Then the vector equation (\ref{vectoreq}) reduces to the two scalar equations
\begin{equation}
	\label{s2xz2}
	\frac{\partial \delta S^{(2)}_{\bm p,x}(u) }{\partial u} - \delta S^{(2)}_{\bm p,z}(u) = 0\:,
	\quad
	\frac{\partial \delta S^{(2)}_{\bm p,z}(u) }{\partial u} + \delta S^{(2)}_{\bm p,x}(u) = -  \delta S^{(1)}_{\bm p,x} \:,
\end{equation}
where a new variable $u = \beta_{yx} p_x^2/e \mathcal{E} \hbar$ is introduced instead of $p_x$. 
Taking into account that (i)~$\delta S^{(1)}_{\bm p,x}$ is independent of $p_x$ in the passive region and (ii)~$\delta \bm S^{(2)}_{\bm p} =0$ at $\beta_{yx}=0$, 
we obtain for the $x$ and $z$ spin density components
\begin{equation}
\label{SxSz}
	\delta S^{(2)}_{\bm p,x}=-\delta S^{(1)}_{\bm p,x}\left(1-\cos{u}\right), 
	\quad
	\delta S^{(2)}_{\bm p,z}=-\delta S_{\bm p,x}^{(1)}\sin{u}. 
\end{equation}
Note that $\delta S_{\bm p,x}^{(1)}$ is odd in $p_y$ and, therefore, the both spin projections vanish after the summation over $\bm p$. As before, the substitution of $\delta S^{(2)}_{\bm p,z}$ into Eq.~(\ref{Gsy}) yields the generation of the spin $y$-component which is nonzero because the product $ \Omega_{\bm p,x} \delta S^{(2)}_{\bm p,z}$ is an even function of $p_y$. Finding the generation rate $G_s/N$ and myltiplying it by $\tau_s$ we get
\begin{equation} \label{synew}
	s_y = - {1 \over 4 \omega_0 \tau} {{\rm Si}(\xi)\over \xi} {\rm sign} (\beta_{yx})\:,
\end{equation}
where $\xi = \sqrt{ \vert \Omega_{{\bm p}_0,y} \vert t_{\rm tr} /\pi} = p_0\sqrt{2|\beta_{yx}|/\pi\hbar e \mathcal{E}}$ and ${\rm Si}$ stands for the Fresnel sine integral.
For small values of $\xi$ the ratio ${\rm Si}(\xi) / \xi$ is approximated by $\pi \xi^2/6$ and Eq.~(\ref{synew}) turns into Eq.~\eqref{s_fin}.

\begin{figure}[t]
\includegraphics[width=0.8\linewidth]{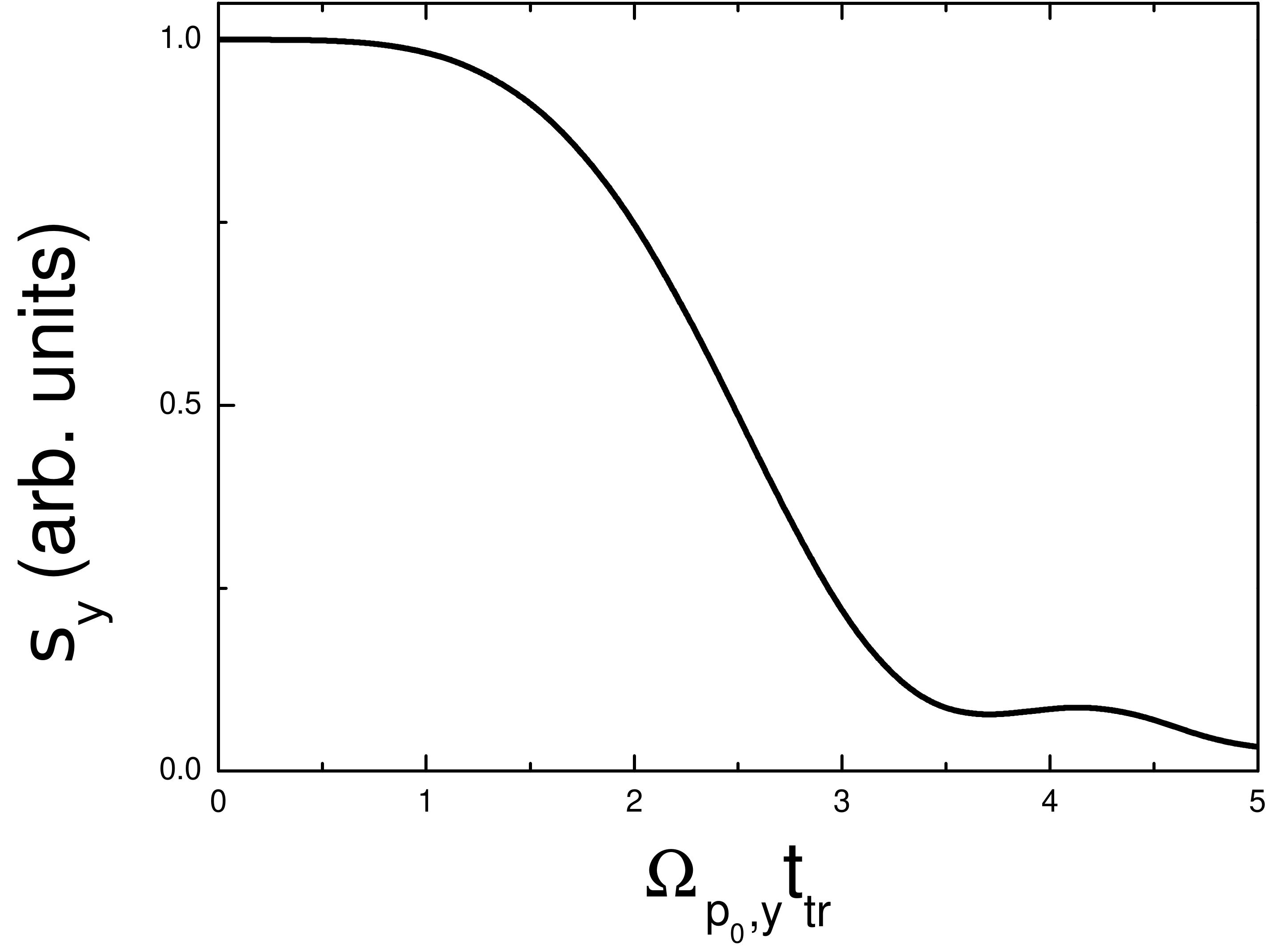}
\caption{Dependence of the induced spin on the precession frequency. Values of $s_y$ are related to the reference spin $s_y^0 = -m \beta_{yx} / (6e\mathcal{E}\tau)$.}
\label{fig:strong}
\end{figure}

To demonstrate the effect of increasing spin rotation frequency $\Omega_{{\bm p}_0,y}$ in the passive region, in Fig.~\ref{fig:strong} we depicted the spin $s_y$ related to its value $s_y^0 = -m \beta_{yx} / (6e\mathcal{E}\tau)$ found at small values of this frequency. One can see that the electrically induced spin is depolarized for $\Omega_{{\bm p}_0,y} t_{\rm tr}> \pi/2$ because the increasing spin-precession rate of the accelerating electron results in a remarkable twisting of the spin similarly to the Hanle effect. The interplay of electric-field effects on the travel time $t_{\rm tr}=p_0/e\mathcal{E}$ and the reference spin value $s_y^0$ results in a maximum of $s_y$ as a function of $\mathcal{E}$, as presented by the curve~3 in Fig.~\ref{fig_spin_vs_E}.

\section{Discussion and Conclusion}
\label{Sec:Disc}

The spin-dependent linear-in-${\bm p}$ Hamiltonian (\ref{H_SO}) arises due to both the Bulk- and Structure-Inversion Asymmetries, in abbreviated form BIA and SIA, respectively~\cite{winkler,Ivchenkobook}.  The BIA contribution follows from the cubic-in-${\bm p}$ Hamiltonian averaged along the quantization axis $z$, i.e., replacing $p_z$ and $p_z^2$ by the average values $\langle p_z \rangle = 0$ and $\langle p_z^2 \rangle \neq 0$. For small electron energies, the remaining cubic contribution to the Hamiltonian
$H_{\rm so}^{(3)}(\bm p) = \gamma (\sigma_{x_0} p_{x_0}p_{y_0}^2 - \sigma_{y_0} p_{y_0}p_{x_0}^2)$
can be ignored as compared to spin-dependent effects governed by the linear-${\bm p}$ Hamiltonian. Here $x_0$ and $y_0$ are the crystallographic axes [100] and [010]. In the streaming regime, however, the electron energy ranges between the conduction-band bottom and the optical-phonon energy and is not small. The allowance for the cubic term results in the following generalized equations for the spin relaxation time and current-induced spin polarization
\begin{equation}
{1\over \tau_{sy}} = 2 \tau
\left[ \frac{p_0}{\hbar} \left(\beta_{xy} - {\gamma p_0^2\over 6}\right) \right]^2 \:,
\quad
s_y = \frac{s_y(\gamma=0)}{(1 - \lambda/6)^2} \left( 1 + \sum\limits_{n=1}^3 c_n \lambda^n \right)\:,
\end{equation}
where $s_y(\gamma=0)$ is the spin polarization in the absence the cubic spin-orbit splitting, see Eq.~(\ref{s_fin}), $\lambda = \gamma p_0^2/\beta_{xy}$,
\[
c_1 = \frac{3}{10} (1 - \mu)\:,\: c_2 = \frac{1}{28} (1 - 3 \mu)\:,\: c_3= -\frac{1}{72}\:,
\]
and $\mu = \beta_{xy}/\beta_{yx}$. If the BIA contribution to the parameters $\beta_{xy}, \beta_{yx}$ prevails they can be estimated by $- \gamma \left< p_z^2\right> \approx - \gamma (\pi \hbar/a)^2$, where $a$ is the well thickness. In this case the allowance for cubic-term contribution gives for $a=100$~\AA \, a decrease by 22\% in $\tau_{sy}$ and by 25\%  for $s_y$.

The mechanism of current-to-spin effect considered up to now is based on the spin-orbit splitting of the 2D conduction subband and the spin-independent electron-phonon coupling: the spin generation rate $G_s$ of the $s_y$ component is proportional to $\beta_{xy}^2 \beta_{yx}$, the spin relaxation rate $\tau^{-1}_{sy}$ of this component is proportional to $\beta_{xy}^2$ leading to $s_y \propto \beta_{yx}$. We have also analyzed another mechanism due to (i) the $\beta_{yx} \sigma_y p_x$ spin-orbit coupling and (ii) electron spin-flips in the LO-phonon emission. In this mechanism,  $G_s \propto \beta_{yx} \lambda_{\rm sf}^2$ and $\tau_{sy} \propto \lambda_{\rm sf}^{-2}$ where $\lambda_{\rm sf}$ is the spin-dependent electron--optical-phonon coupling constant.
In bulk semiconductors, for the Fr\"ohlich polar-optical electron-phonon interaction and the Elliott-Yafet mechanism of spin relaxation, the matrix element of the spin-flip phonon emission in the process $({\bm p}',s') \to ({\bm p},s) + \hbar \omega_0$
can be written in the 2$\times$2 matrix form as~\cite{Pikus}
\begin{equation}
\hat{M}_{{\bm p} {\bm p}^{\prime}}= \frac{{\rm i}\hbar C_{\rm F}}{\sqrt{V} \, |{\bm p} - {\bm p}'|} 
\left[ 1 + {\rm i} \lambda_{\rm sf} {\bm \sigma} \cdot ({\bm p} \times {\bm p}') \right]\:,
\end{equation}
where the constant $\lambda_{\rm sf} = B(2A + B)$,
\[
A = {P\over 3} \frac{3 E_g + 2 \Delta}{E_g(E_g + \Delta)} \:,\qquad B = -{P\over 3} \frac{\Delta}{E_g(E_g + \Delta)}\:,
\]
$P$ is the Kane matrix element, $E_g$ is the energy gap, and $\Delta$ is the spin-orbit splitting of the valence band.
In 2D systems, the matrix element of the electron-phonon interaction acquires the form~\cite{IOPP}
\[
	\hat{M}_{{\bm p} {\bm p}^{\prime}}(q_z)= 	
M_{{\bm p} {\bm p}^{\prime}}(q_z)  \left\{ 1 - {\rm i} \lambda_{\rm sf} {\hbar q_z \over 2} [\sigma_x (p_y + p_y')-\sigma_y (p_x + p_x')] \right\}\:,
\]
where the spin-independent matrix element $M_{{\bm p} {\bm p}^{\prime}}(q_z)$ is defined in Eq.~(\ref{froehlich2}).
Using this equation we have estimated the contribution of spin-flips to both the spin relaxation rate $\tau^{-1}_{sy}$ and the spin generation rate $G_s$ and found that they are much smaller as compared with those calculated in Section~\ref{Sec:el_spin_orient}.

All along the paper we fixed the attention on the behavior of the $s_y$ spin component generated by the electron acceleration in the $x$ direction. The relaxation of this component in the streaming regime is described by the time $\tau_{sy}$ determined by Eq.~(\ref{tauxyz}). In addition, this regime offers intriguing behavior of the nonequilibrium spin polarization perpendicular to $y$ due to precession caused by the electric field~\cite{Kalevich_Korenev}. Indeed, let at $t=0$ a portion of spin-polarized photoelectrons be injected into the conduction-subband bottom by a short optical circularly-polarized pulse. If the electrons are initially polarized along the $z$ axis, only the LO-phonon emission is taken into account in the collision integral and the small precession frequency component $\Omega_{{\bm p}x}$ is neglected, then the time-dependent spin-distribution function of the photoelectrons has the form
\[
\bm S_{\bm p}(t) =2 \tilde{f}_{\bm p} {\bm s}(t) p_0 \delta \left(p_x - e\mathcal{E}t_{\rm tr}\left\{ {t \over t_{\rm tr}}\right\}\right)\:.
\]
Here $\tilde{f}_{\bm p}$ is the function (\ref{spin_independ}) normalized to the photoelectron density $\tilde{N}$, 
\[
s_z(t) =s_z^{(0)} \cos{\Phi(t)}\:,\:	s_x(t) = s_z^{(0)} \sin{\Phi(t)}\:,
\]
$s_z^{(0)}$ is the initial spin, and the phase $\Phi$ changes in time according to 
\begin{equation} \label{phi}
\Phi(t)=  R \biggl( \left[ {t \over t_{\rm tr}}\right] + 
\left\{ {t \over t_{\rm tr}}\right\}^2 \biggr)\:,
\end{equation} 
with $R$ being ${e\mathcal{E}\beta_{yx}t_{\rm tr}^2/\hbar = \beta_{yx}p_0^2/e\mathcal{E} \hbar}$ and the symbols $[x]$, $\{x\}$ denoting, respectively, the integer and fractional parts of $x$. If $R$ is a rational part of $2 \pi$ the functions $s_z(t), s_x(t)$ are periodic. For example, if $R = 2 \pi/n$ with an integer $n$ the period equals to $n t_{\rm tr}$. For irrational values of $R/2 \pi$, the variation $s_{z,x}(t)$ is aperiodic. Allowance for other mechanisms of electron scattering and nonzero $\Omega_{{\bm p}x}$ leads to decay of the oscillation amplitude. In the presence of a transverse magnetic field ${\bm B} \parallel y$, the contribution
\[
\Omega_{\bm B} t  = \Omega_{\bm B} t_{\rm tr} \biggl( \left[ {t \over t_{\rm tr}}\right] + 
\left\{ {t \over t_{\rm tr}}\right\} \biggr)
\]
should be added to (\ref{phi}), where $\Omega_{\bm B}$ is the Larmor angular frequency.

In conclusion, the spin-dependent streaming theory is developed for semiconductor 2D systems. The particular attention is payed to electron spin polarization induced by a dc electric current in a quantum well. The mechanism of spin generation by the current presumes spin-orbit splitting of the electron quantum-confined subbands. The nonzero average spin polarization arises taking into account a spin-dependent quantum correction to the collision integral and the following precession of this spin correction around the effective magnetic field related to the spin-orbit Hamiltonian. The main finding is a considerable enhancement of the spin orientation in the streaming regime resulting in a few percent spin polarization. In case of the high spin-orbit coupling strength, the spin polarization behaves non-monotonically as a function of the electric field. In addition to effects of the linear spin-orbit coupling, we have analyzed the role  of spin-orbit Hamiltonian cubic in the electron momentum as well as Elliott-Yafet spin-flip processes. It is also shown that the transient spin dynamics in the streaming regime is very reach and presents many interesting phenomena including periodic oscillations of the photocreated spin in time at particular values of the electric field.

\ack{We thank V.A.~Kozlov and V.L. Korenev for discussions.
The work was supported by Samsung Electronics, RFBR and EU programmes SPANGL4Q and POLAPHEN.}


\begin{thebibliography}{36}
\bibitem{inj1} A.G. Aronov and G.E. Pikus, Fiz. Tekh. Poluprovodn.
{\bf 10}, 1177 (1976) [Sov. Phys. Semicond. {\bf 10}, 698 (1976)].

\bibitem{inj2} D.D. Awschalom and M.E. Flatte, Nature Phys. {\bf 3}, 153 (2007).

\bibitem{inj3} Kun-Rok Jeon, Byoung-Chul Min, Young-Hun Jo, Hun-Sung Lee, Il-Jae Shin, Chang-Yup Park, Seung-Young Park, and Sung-Chul Shin, Phys. Rev. B {\bf 84}, 165315 (2011).

\bibitem{SG_EL}  \textit{Spin Physics in Semiconductors}, ed. M.I. Dyakonov (Springer, 2008).

\bibitem{Rashba_term}  E.I. Rashba, Fiz. Tverd. Tela (Leningrad) \textbf{2}, 1224 (1960) [Sov. Phys. Solid State \textbf{2}, 1109 (1960)]; Y.A. Bychkov and E.I. Rashba, J. Phys. C \textbf{17}, 6039 (1984).

\bibitem{IvchPikus} E.L. Ivchenko and G.E. Pikus,  Pis'ma Zh. Eksp. Teor.
Fiz. {\bf 27}, 640 (1978) [JETP Lett. {\bf 27}, 604 (1978)].

\bibitem{cpge2}L.E. Vorob'ev, E.L.~Ivchenko, G.E.~Pikus, I.I.
Farbshtein, V.A. Shalygin, and A.V. Shturbin, Pis'ma Zh. Eksp. Teor.
Fiz. {\bf 29}, 485 (1979) [JETP Lett. {\bf 29}, 441 (1979)].

\bibitem{cpge3} V.A. Shalygin, A.N. Sofronov, L.E. Vorob'ev, and I.I.
Farbshtein, Fiz. Tverd. Tela {\bf 54}, 2237 (2012) [Phys. Solid State {\bf 54}, 2362 (2012)].

\bibitem{Kato04} Y.K. Kato, R.C. Myers, A.C. Gossard, and D.D.~Awschalom,
Phys. Rev. Lett. \textbf{ 93}, 176601 (2004).

\bibitem{Vasko} F.T.~Vasko and N.A. Prima, Fiz. Tverd. Tela {\bf
21}, 1734 (1979) [Sov. Phys. Solid State {\bf 21}, 994 (1979)].

\bibitem{Aronov} A.G. Aronov and Yu.B. Lyanda-Geller,
Pis'ma Zh. Eksp. Teor. Fiz. \textbf{ 50}, 398 (1989)
[JETP Lett. \textbf{50}, 431 (1989)].

\bibitem{Edelstein} V.M. Edelstein, Solid State Commun. \textbf{ 73},
233 (1990).

\bibitem{Aronov91} A.G. Aronov, Yu.B. Lyanda-Geller, and G.E. Pikus,
Zh Eksp. Teor. Fiz. \textbf{ 100}, 973 (1991)
[Sov. Phys. JETP \textbf{ 73}, 537 (1991)].

\bibitem{Ganichev06}  S.D. Ganichev, S.N. Danilov, Petra Schneider, V.V. Bel'kov, L.E. Golub,
W. Wegscheider, D.Weiss, and W. Prettl, cond-mat/0403641 (2004), see also J. Magn. Magn. Mater. \textbf{300}, 127 (2006).

\bibitem{Silov2004} A.Yu. Silov, P.A. Blajnov,  J.H. Wolter,
R.~Hey, K.H.~Ploog, and  N.S.~Averkiev,
\textit{Appl. Phys. Lett.} \textbf{ 85}, 5929 (2004).

\bibitem{spinorient_PRB} L.E. Golub and E.L. Ivchenko, Phys. Rev. B \textbf{84}, 115303 (2011).

\bibitem{Hosur} P. Hosur, Phys. Rev. B {\bf 83}, 035309 (2011).

\bibitem{Japan_TI} 
T. Misawa, T. Yokoyama, and S. Murakami,
Phys. Rev. B \textbf{84}, 165407 (2011).

\bibitem{Rashba} E.I. Rashba, Phys. Rev. B {\bf 86}, 125319 (2012).

\bibitem{Streaming_bulk_review} E. Gornik and A.A. Andronov (eds.), Opt. Quantum Electron. \textbf{23}, s111 (1991).

\bibitem{Andronov80} A.A. Andronov, V.A. Valov, V.A. Kozlov and L.S. Mazov, Solid State Commun. \textbf{36}, 603 (1980).

\bibitem{Dmitriev_Tsendin} 
A.P. Dmitriev and L.D. Tsendin, Sov. Phys. JETP \textbf{54}, 1071 (1981) [Zh. Eksp. Teor. Fiz. \textbf{81}, 2032 (1981)].

\bibitem{ridley} B.K. Ridley, J. Phys. C: Solid State Phys. {\bf 17}, 5357 (1984).

\bibitem{Streaming_theory_1a} 
V.V. Korotyeyev, V.A. Kochelap, K.W. Kim, and D.L. Woolard, Appl. Phys. Lett. \textbf{82}, 2643 (2003).

\bibitem{Streaming_theory_1b} 
K.W. Kim, V.V. Korotyeyev, V.A. Kochelap,  A.A. Klimov, and D.L. Woolard,  J. Appl. Phys. \textbf{96} 6488 (2004).
\bibitem{Streaming_theory_2} 
J.T. L\"{u} and J.C. Cao, Semicond. Sci. Technol. \textbf{20},  829 (2005).

\bibitem{cardona} P.Y. Yu and M. Cardona, \textit{Fundamentals of Semiconductors: Physics and Materials
Properties} (Springer, New York, 2001), 3rd ed.

\bibitem{IvAnov} Yu.L. Ivanov, G.V. Churakov, V.M. Ustinov, A.E. Zhukov, and A.Yu. Egorov, Semiconductors \textbf{29}, 1702 (1997).

\bibitem{ILGP89} E.L. Ivchenko, Yu. B. Lyanda-Geller, and G.E. Pikus, Pis'ma Zh. Eksp. Teor. Fiz \textbf{50}, 156 (1989) [JETP Lett. \textbf{50}, 175 (1989)].
\bibitem{ILGP90} E.L. Ivchenko, Yu. B. Lyanda-Geller, and G.E. Pikus, Zh. Eksp. Teor. Fiz \textbf{98}, 989 (1990) [Sov. Phys. JETP \textbf{71}, 550 (1990)].
\bibitem{noci} S. Gantsevich, V. Gurevich, and R. Katilius,
    Riv. Nuovo Cimento {\bf 2}, 1 (1979).

\bibitem{fabian} J. Fabian, A. Matos-Abiaguea, C. Ertlera, P. Stano, and I. \v{Z}uti\'c, Acta Physica Slovaca \textbf{57}, 565 (2007). 

\bibitem{winkler} R. Winkler, {\it Spin-Orbit Coupling Effects in Two-Dimensional Electron and Hole Systems} (Springer-Verlag, Berlin, Heidelberg, 2003).

\bibitem{Ivchenkobook} E.L.~Ivchenko, \textit{ Optical Spectroscopy of Semiconductor Nanostructures} (Alpha Science Int., Harrow, UK, 2005).

\bibitem{Pikus} G.E. Pikus and A.N. Titkov, in \textit{Optical Orientation} (Amsterdam, North-Holland, 1984).
\bibitem{IOPP} N.S. Averkiev, L.E. Golub and M. Willander, J. Phys.: Condens. Matter \textbf{14}, R271 (2002).

\bibitem{Kalevich_Korenev} V.K. Kalevich, V.L. Korenev, Pis'ma Zh. Eksp. Teor. Fiz \textbf{52}, 859 (1990) [JETP Lett. \textbf{52}, 230 (1990)].
\end{thebibliography}
\end{document}